\documentclass[preprint,floatfix,11pt]{revtex4-1}
\usepackage{color,graphicx}
\usepackage{bm}
\usepackage{amsmath}
\usepackage{amssymb}

\newcommand{\be}{\begin{eqnarray}}
\newcommand{\ee}{\end{eqnarray}}

\begin{document}

\title{Quantum critical response function in 
quasi-two dimensional itinerant antiferromagnets}
\author{C. M. Varma and Lijun Zhu}
\affiliation{Department of Physics and Astronomy, University of California, Riverside CA 92521, USA}

\author{Almut Schr\"oder}
\affiliation{Department of Physics, Kent State University, Kent, OH 44242}

\pacs{}
\date\today
\begin{abstract}
We re-examine the experimental results for the magnetic response function $\chi''({\bf q}, E, T)$, for ${\bf q}$ around the anti-ferromagnetic vectors ${\bf Q}$, in the quantum-critical region, obtained by inelastic neutron scattering, on an Fe-based superconductor, and on a heavy Fermion compound. The motivation is to compare the results with a recent theory, which shows that the fluctuations in a generic anti-ferromagnetic model for itinerant fermions map to those in the universality class of the dissipative quantum-XY model. The quantum-critical fluctuations in this model,
in a range of parameters, are given by the correlations of spatial and of temporal topological defects. The theory predicts a $\chi''({\bf q}, E, T)$ (i) which is a separable function of $({\bf q -Q})$ and of ($E$,$T$), (ii) at crticality, the energy dependent part is $\propto \tanh (E/2T)$ below a cut-off energy, (iii) the correlation time departs from its infinite value at criticality on the disordered side by an essential singularity, and (iv) the correlation length depends logarithmically on the correlation time, so that the dynamical critical exponent  $z$ is $\infty$ . The limited existing experimental results are found to be consistent with the first two unusual predictions from which the linear dependence of the resistivity on T and the $T \ln T$ dependence of the entropy also follow. More experiments are suggested, especially to test the theory of variations on the correlation time and length on the departure from criticality.

\end{abstract}
\maketitle

\section{Introduction}
Soon after the discovery that in the normal metallic region for dopings near those for the high superconducting transitions in cuprates have transport and thermodynamic properties \cite{*[{}]  [{ as well as the preceding four volumes of this series.}]GinsburgV} unlike those expected of a Fermi-liquid, it was discovered that the heavy-Fermion compounds, for compositions near where their anti-ferromagnetic (AFM) transition temperatures $T_N \to 0$, also have similar anomalies \cite{HvL07}. For example, in both cases the resistivity has a temperature dependence proportional to $T$ at low temperatures and the entropy or thermopower (entropy per carrier) has a contribution proportional to $T \ln T$.  More recently, the AFM quantum-critical region in the Fe based superconductors also show the same anomalies in the resistivity and the thermo-power (see, e.g., Refs.\onlinecite{Zheng,Analytis}).

The theoretical study of quantum-criticality in AFM's began long before these experiments came along, starting with the work of Moriya, and Hertz and others\cite{Moriya,Hertz,Millis,Beal-monod}. These are extensions of the theory of classical dynamical critical phenomena \cite{Hohenberg} to the quantum-regime and may be called renormalized spin-wave theories. Since the above experiments, an enormous amount of theoretical work in further developing theories based on similar ideas has been done \cite{Millis}.  The results from such theories appear to correspond to experiments in AFM's in which the fluctuations are 3-dimensional (3d) \cite{CeCu2Si2Neutrons,Nishiyama2013} but not for those in which they are two-dimensional (2d). In the past few years, S-S. Lee and collaborators \cite{SSLee13,*SSLee14a,*SSLee14b} have shown that such theories are not controlled in 2d. This has been followed by significant theoretical work which seeks to approach the actual 2d problem by expanding about limits \cite{SSLee13,*SSLee14a,*SSLee14b,Metlitski10a,*Metlitski10b,EfetovPepin} in which the theory is controlled, for example about a Fermi-surface in 3d while the spin-fluctuations are 2d. Another set of ideas for the heavy-fermion problem relies on the fact that an isolated Kondo impurity in a metal has singular properties near the criticality of the host \cite{Si01,*Si03} and by the approximation that the heavy-fermion metal may be regarded as a self-consistent set of periodic Kondo impurities using methods such as the Dynamical mean-Field Theory. Yet another seeks to understand the results within the renormalized spin-wave framework by invoking a phenomenological renormalization of the spatial correlation length \cite{Woelfle11,*Abrahams12}. More exotic ideas have also been proposed \cite{Coleman01}.

A radically different solution to the 2d AFM problem has recently been proposed \cite{Varma15}. It starts by showing that the model for criticality of itinerant AFM's is the dissipative XY model (with appropriate lattice anisotropy). This is true for the AFM with planar order, either about a commensurate or incommensurate wave vector, or Ising order about an incommensurate wave vector. The quantum-dissipative XY model \cite{Chakravarty86} in 2d in such a region has been transformed \cite{Aji07,*Aji09,*Aji10} to a model in which the critical properties (on the disordered side) are determined by topological excitations - 2d vortices as in the classical 2d-XY model and a new variety of topological excitations, called warps which are instantons of bound monopole-anti-monopoles with net charge 0. 
Two dimensionless parameters specify a line along which the quantum-critical point occurs in the model, which may be taken to be $\sqrt{KK_{\tau}}$ and $\alpha$. $K$ is the Josephson coupling energy of the phase variables, $K_{\tau}^{-1}$ is the magnitude of the kinetic energy parameter, and $\alpha$ is a dimensionless parameter for the dissipation due to decay of phase fluctuations\cite{Sudbo12,Zhu15}.
The quantum-criticality in this space, from the disordered state to the ordered state, appears in quantum Monte-Carlo calculations to have either a dynamical critical exponent $z = 1$ or $z = \infty$, depending on the critical value of the parameters\cite{ZhuCV}. Only the $z=\infty$ case, which occupies the bulk of the parameter space in the $\sqrt{KK_{\tau} }- \alpha$ plane is summarized below, as this alone provides results in correspondence which bear relation to the theory.

A stringent test of the theory requires measurements of the absolute magnitude of $\chi''({\bf q}, E, T)$ in the fluctuation regime over a range of ${\bf q}$ and spanning frequencies from well above to well below $T$. The thermodynamic and transport properties can  usually be derived from $\chi''({\bf q}, E, T)$. The purpose of this paper is to present the limited available existing quantitative experimental results for this function, which are available in the range of ${\bf q}$, $E$, and $T$ to compare and check whether there is a correspondence between the predictions of the theory. 

\section{Correlation Functions}

The correlations of the dissipative quantum 2d-XY model have been recently investigated in analytical calculations \cite{Aji07,*Aji09,*Aji10}. These have been checked and extended by quantum Monte-Carlo calculations \cite{Zhu15}.  These calculations confirm that the transitions are driven by vortex and/or warp binding and not by (renormalized) spin-waves which only serve to generate effective interactions among the topological charges. The calculated correlation functions as a function of distance $r$ and imaginary time $\tau$ on the disordered side by
\be
\label{chi-r}
\chi(r, \tau) &\approx & \chi_0 \frac{1}{\tau} e^{-\sqrt{\tau/\xi_{\tau}}} \ln \big(\frac{r_c}{r}\big) e^{-r/\xi_r} e^{i{\bf Q}.{\bf r}}, 
\\ 
\label{xi-tr}
\xi_{\tau} &=& |\tau_c|~ e^{\sqrt{{\frac{p_c}{p_c-p}}}}; ~~ \xi_r/r_c \approx \ln (\xi_{\tau}/|\tau_c|).
\ee
Here $\tau$ is the imaginary time, periodic in $1/(k_B T)$ ($k_B=1$ in this paper), which has a lower cut-off $|\tau_c| \approx (K K_{\tau})^{-1/2}$ and $r_c$ is the short distance spatial cut-off, of the order of a lattice spacing.  $p$ is the parameter which tunes the transition by tuning $KK_{\tau}$ or $\alpha$, and $p_c$ is the critical value of $p$  at which the transition occurs at $T \to 0$. $\chi_0$ together with the high frequency cut-off serves to give the integrated magnetic fluctuations through the sum-rule on them. 

To compare with experiments, it is necessary to Fourier transform the correlation function to momentum $({\bf q})$ and frequency ($E$) variables. The Fourier transform to frequency space can be reduced to an integral which can only be evaluated numerically. The results and the fits to it to a functional form for the imaginary part are given in Ref. (\onlinecite{Zhu15}). We quote this result: 
\be
\label{chi-tr}
\chi''({\bf q}, E, T) &=& - \chi_0  \tanh\left(\frac{E}{2T}\right) \frac{1}{|{\bf q- Q}|^2 + \kappa_q^{2}}{\mathcal{F}}_{\ell}\left(\frac{ T}{\kappa_E} \right){\mathcal{F}}_u\left(\frac{E}{E_c}\right) , 
\ee 
 ${\mathcal{F}}_{\ell}\left(\frac{ T}{\kappa_E} \right)$ serves as an infra-red cut-off function due to deviations from criticality, where $\kappa_E \equiv \xi_{\tau}^{-1}$, replaces $T$ as the infrared energy scale for critical fluctuations. Note that $\xi_{\tau}^{-1}$ increases extremely slowly, as an essential singularity (see Eq. (\ref{xi-tr})) from 0 on deviation from criticality. 
${\mathcal{F}}\left(\frac{ T}{\kappa_E} \right) \to 1$ at criticality when $\kappa_E \to 0$, so that the response is simply proportional to $\tanh(E/2T)$ with an ultra-violet cut-off at $E_c \equiv |\tau_c|^{-1}$. For finite $\kappa_E$, a fit to the numerical results gives,
\be
{\mathcal{F}}_{\ell}\left(\frac{ T}{\kappa_E} \right) & \approx & \frac{1}{\left(1 +  \sqrt{ \kappa_E/2\pi T}\right)^2}, ~\text{for}~ E/ T \ll 1; \\ \nonumber
& \approx & \frac 14 \left(1+ 3e^{-\sqrt{\kappa_E/ T}}\right) ~\text{for}~E_c/T \gg E/T \gg 1.
\ee
$\kappa_q = \xi_r^{-1}$. ${\mathcal{F}}_u\left(E/E_c\right)$ serves as an ultraviolet cutoff function  ${\mathcal{F}}_u(0) =1,{\mathcal{F}}_u(\infty) = 0.$  In experiments, the ultra-violet cut-off  scale may come from physics not in the effective low-energy model, as for example the Fermi-energy, if it is smaller than $ |\tau_c|^{-1}$.  We will be confronted with this situation below in the measurements in the heavy-fermion compound CeCu$_{6-x}$Au$_x$. 

The most striking result in Eq. (\ref{chi-tr}) is that $ \chi''( {\bf q}, E,T)$ is a separable function of $(E,T$) and of ${\bf q}$. Other remarkable results are that the dimensionless characteristic correlation wave-vector depends logarithmically on the characteristic correlation frequency. This means that the dynamical critical exponent is $z = \infty$. 
Also, Eq. (\ref{chi-r}) for $\xi_{\tau}$ gives that the deviation of the characteristic correlation frequency with departure from criticality is determined by an essential singularity from its value 0 at criticality. In other words, the cross-over from what is usually called the quantum-fluctuation regime, where the fluctuation scale is $T$ to the truly quantum-regime, where the fluctuations are determined by $(p-p_c)/p_c$ is extremely slow. 

The frequency and temperature dependence of these fluctuations at criticality is the same as that proposed phenomenologically \cite{Varma89} to give the marginal Fermi-liquid of fermions in the cuprates. The result for the spatial correlation length are quite different and remove the un-appetizing feature of the phenomenology that the spatial correlation length was independent of deviation from criticality. However, the single particle self-energy and the resistivity calculated due to scattering from the fluctuations given by Eq. (\ref{chi-tr}) is also proportional to $T$ in the quantum-critical regime, $T /\kappa_E \gtrsim 1$ (see supp. sec, in Ref. [\onlinecite{Varma15}]) and the entropy is proportional to $T \ln T$. A microscopic origin \cite{Aji07,*Aji09,*Aji10}, quite different from AFM, has been found for criticality in the relevant range of cuprates, whose fluctuations also map to the 2d-XY model.

The most stringent test of the theory is through the measurement of $ \chi( {\bf q}, E,T)$, from which most other properties can be derived. These are very difficult measurements which also require large crystals. We compare here the results with the existing measurements in a Fe-based compound BaFe$_{1.85}$Co$_{0.15}$As$_2$, and a heavy-Fermion compound CeCu$_{6-x}$Au$_x$. These are the only measurements that we are aware of which are done in the frequency, momentum and temperature region of interest to test theories of AFM quantum-criticality. We will also recall some old results in a cuprate compound near the low doping where its AFM transition temperature appears $\to 0$ but in which, due to the disorder, a spin-glass type phase sets in at a finite temperature.

\section{$\chi''({\bf q}, E, T)$ in B\lowercase{a}F\lowercase{e}$_{1.85}$C\lowercase{o}$_{0.15}$A\lowercase{s}$_2$}

\begin{figure}[tbh]
\centering
\includegraphics[width=0.8\columnwidth]{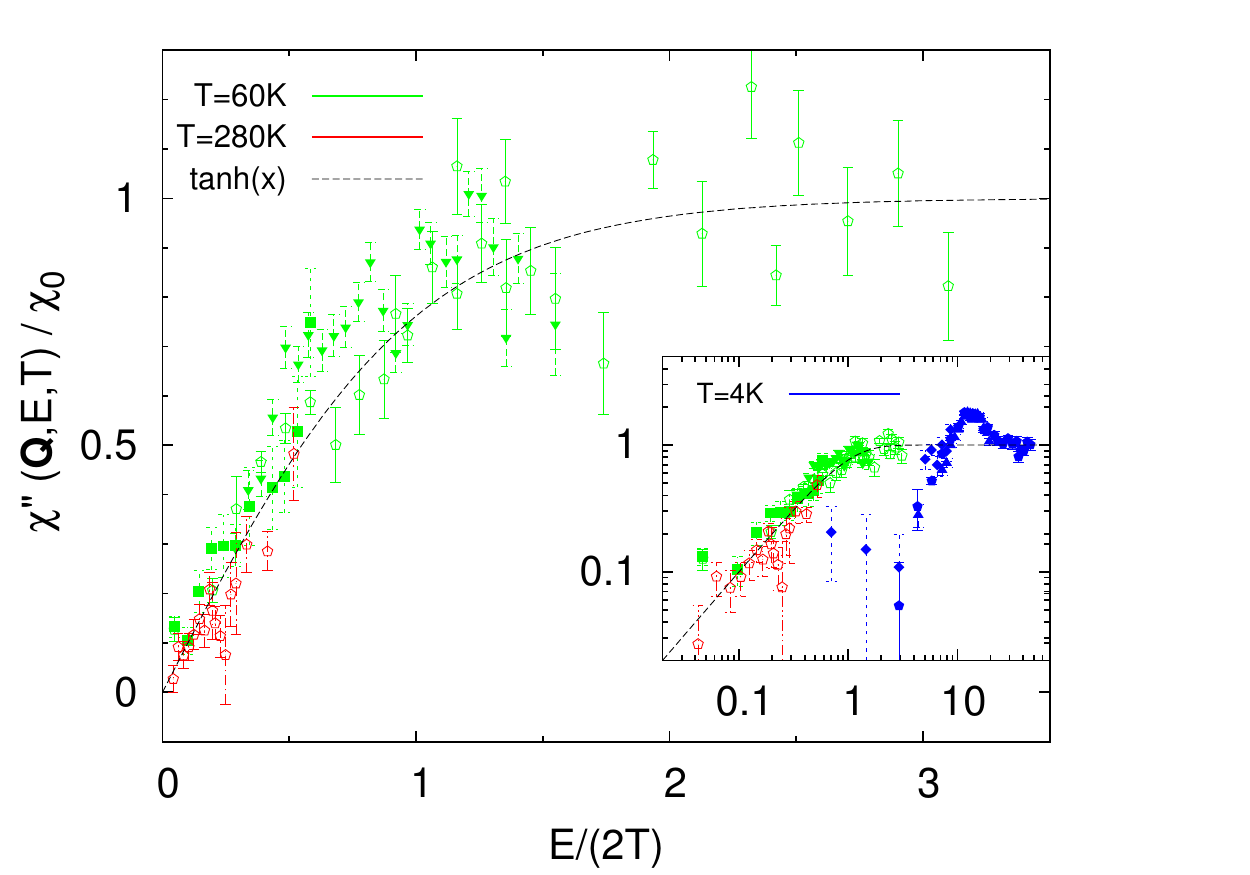}
\caption{$\chi '' ({\bf Q}, E, T)$ as functions of $E/(2T)$ with ${\bf Q}$=(0.5,0.5,1) at $T$=60K and 280K for BaFe$_{1.85}$Co$_{0.15}$As$_2$.  
The data is taken from Ref.~[\onlinecite{Inosov10}]. The inset shows the same data on a log-log scale, so as to also include the $T=4K$ data (well below the superconducting transition temperature), showing that for $E \gtrsim$ 20 meV and up to 50 meV, the data is quantitatively is the same at the higher temperatures for $E/2T \gg 1$, as given by the theory.}
\label{fig:chiet}
\end{figure}

BaFe$_{1.85}$Co$_{0.15}$As$_2$ has a putative Antiferromagnetic quantum critical point of the planar variety very close to the above composition, as we will show from the measurement of the spatial correlation length \cite{footnote-afm}. The fluctuations in the range of frequencies and temperature measured are about ${\bf Q}=(1/2,1/2, 1)$ and equivalent points. The compound is strongly two-dimensional and the fluctuations are studied only in the $(h,k)$- plane. The
superconducting transition temperature of 25 K cuts off the low frequency part of the critical fluctuations and gives a peak in the frequency dependence at near twice the superconducting gap, about 10 meV,  and half-width of peak is about 4 meV. The ${\bf q}$-dependence has been measured \cite{Inosov10} at three different frequencies, 3, 9.5 and 16 meV and at two different temperatures, one at 4 K, well below the superconducting transition temperature, and the other at 60 K. The frequency dependence at the critical vector for many frequencies from about 0.5 meV to 60 meV has been measured at these two temperatures and at 280 K. We have to discard the 4 K data below about 15 meV, because of the superconductivity induced low energy features below about twice the superconducting gap, as further discussed in the caption to Fig. (\ref{fig:chiet}). A great virtue of the measurements is that the absolute intensity of the magnetic scattering has been measured. We therefore need to normalize all the data presented only once for all frequencies and momenta. 

\begin{figure}[tbh]
\centering
\includegraphics[width=0.8\columnwidth]{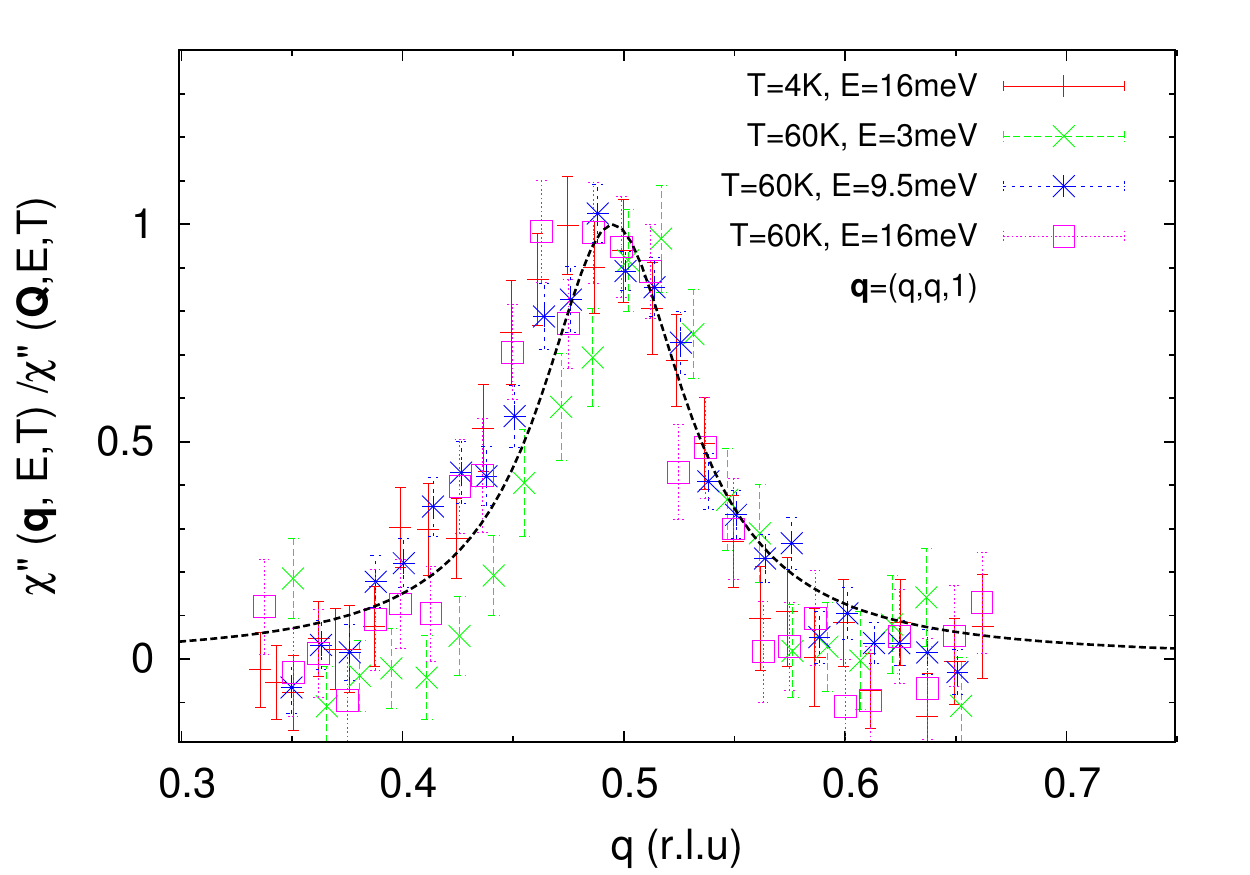}
\caption{$\chi '' ({\bf q}, E, T)$ as functions of $|{\bf q}-{\bf Q}|$ at various fixed temperatures and energies.  Here, the ${\bf q}$-scan is along $(q,q,1)$ and ${\bf Q}= (0.5,0.5,1)$.  A background contribution to $\chi''$ has been subtracted (assuming a linear dependence on $q$ as given in Ref.~[\onlinecite{Inosov10}] and fitted by the Lorentzian $1 /[ (|{\bf q}-{\bf Q}|/\kappa_q)^2 + 1]$  (the curve shown has $\kappa_q$=0.04 r.l.u.).   }
\label{fig:chiq}
\end{figure}

We present the data for the q-dependence at the various indicated frequencies and temperatures  in Fig. (\ref{fig:chiq}). As shown, the distribution in $q$ about the maximum fits a Lorentzian, with a width $\kappa_q \approx 0.04  \pm 0.007 \text{ r.l.u.}$, which is frequency and temperature independent to within the error bars, in the range (a factor of 5 in frequency and 15 in temperature) that it has been measured. This is consistent with the theoretical result that the q and the $E, T$-dependence are separable. (The discarded data at 4 K for frequencies below twice the superconducting gap shows a 20\% smaller $\kappa_q$.)

Taking the measured large upper cut-off \cite{Dai-rev-iron} of the fluctuations in this compound of about 200 meV, and using the measured $\kappa_q \approx 0.04$rlu, Eq. (\ref{chi-r}) gives $\kappa_{E} \ll 4 K$, the lowest temperature measured. Therefore, we should compare the frequency dependence with the form expected at criticality, i.e. $\propto \tanh(E/2T)$. Fig. (\ref{fig:chiet}) shows the absolute measurement of $\chi''({\bf Q}, E, T)/\chi''({\bf Q}, E/T \gg 1)$ at the peak of its momentum-dependence at 60 K and 280 K for all values of energy measured. Also shown is the function $\tanh(E/2T)$. In an inset we include the data at 4 K also. Above the superconductivity induced reduction followed by the bump, it also joins on to the data at 60 K and 280 K data in absolute value. This means that once we fix the magnitude $\chi_0$ of Eq. (\ref{chi-tr}), all the frequency and temperature dependent data is consistent with the form $\tanh(E/2T)$.

Two of the principal predictions of the theory are therefore shown to be obeyed. We urge measurements at various other compositions to test the dependence of the correlation lengths on deviations from criticality as well as more detailed $E$ and $T$ measurements.

\section{\text{$\chi''({\bf q}, E$, T) in C\lowercase{e}C\lowercase{u}$_{6-\lowercase{x}}$A\lowercase{u}$_{\lowercase{x}}$}}

\begin{figure}[tbh]
\centering
\includegraphics[width=0.6\columnwidth]{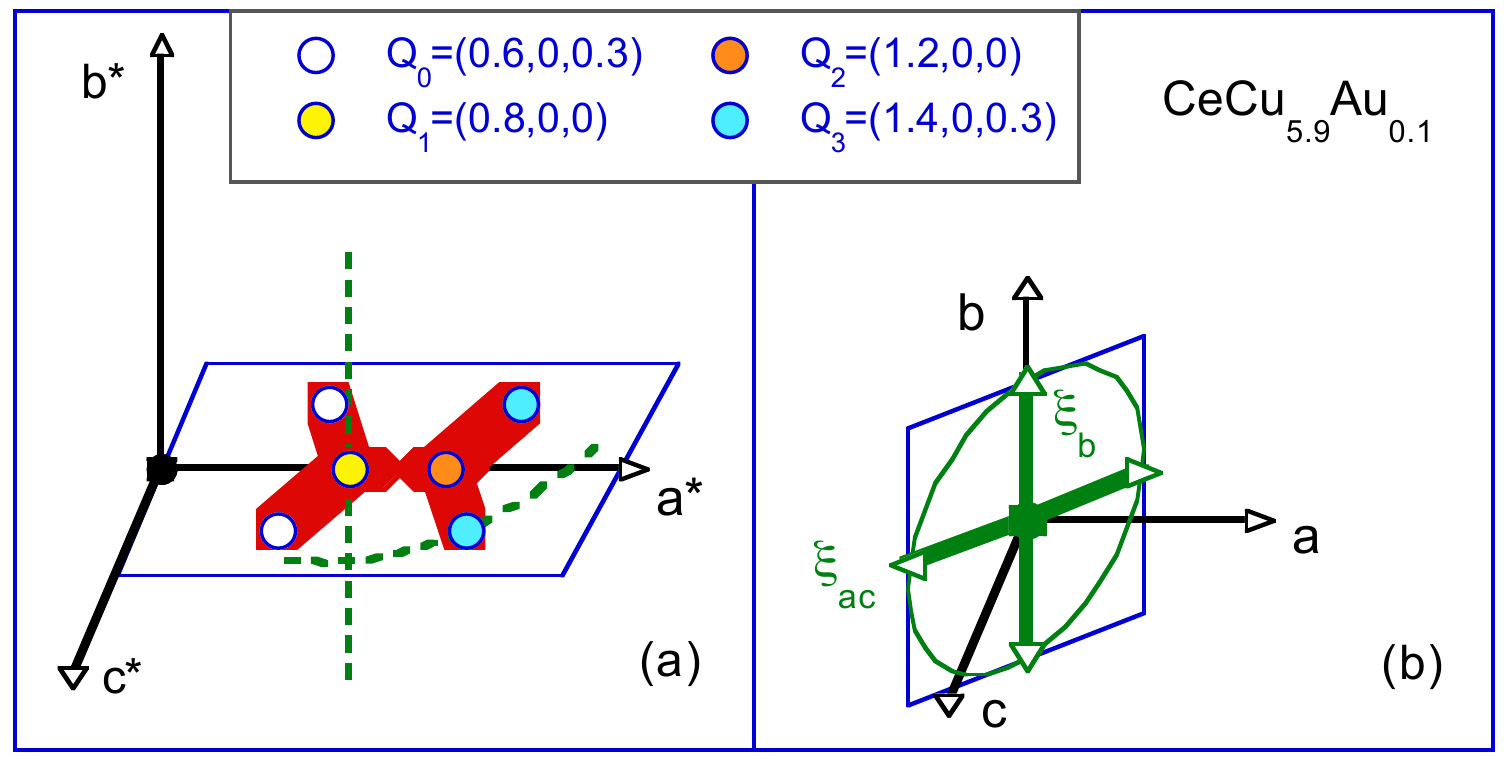}
\caption{(a) The critical wave vectors ${\bf Q}$ of CeCu$_{5.9}$Au$_{0.1}$, shown in reciprocal lattice. The experimental $q$-scan trajectories around ${\bf Q}_1$ and ${\bf Q}_3$ are also shown. (b) The plane of 2d fluctuations is shown in real space. Perpendicular to this plane, $\chi''({\bf q}, E, T)$ is nearly independent of ${\bf q}$, testifying the 2d nature of the fluctuation. Note that the plane of critical fluctuations is not a simple crystallographic plane.}
\label{Fig:cecuau_qorder}
\end{figure}

In CeCu$_{6-x}$Au$_x$, Ising long range AFM order occurs for $x>x_c=0.1$ at an incommensurate vector ${\bf Q_0} \approx (0.6,0,0.3)$ \cite{hvl89} up to $x=0.3$.  The magnetic moments are aligned in the c-direction \cite{hvl89}.  At  the critical concentration $x_c$ the AFM fluctuations are strongest at $ (0.6,0,0.3)$ but also at a wave vector $(0.8,0,0)$ \cite{AS98,OS98}, which only develops short range order for higher $x=0.2$ \cite{hvl89}. Similar response $\chi''({\bf q},E,T)$ for both Q vectors (and equivalent positions) were observed in the measured neutron scattering function for $ x \approx x_c$  along  rods in reciprocal space \cite{OS98,AS00} from $Q_1=(0.8,0,0)$ towards $Q_0=(0.6,0,\pm0.3)$ as sketched in Fig.(\ref{Fig:cecuau_qorder}a). This is consistent with two-dimensional magnetic correlations in real space, perpendicular to the rods, along the b-axis and in the (ac)-plane, as illustrated schematically in Fig. (\ref{Fig:cecuau_qorder}b) for one example.
The neutron scattering data presented here was collected from different spectrometers \cite{AS98,AS00} and calibrated using the incoherent nuclear scattering from the sample.

\begin{figure}[tbh]
\centering
\includegraphics[width=0.6\columnwidth]{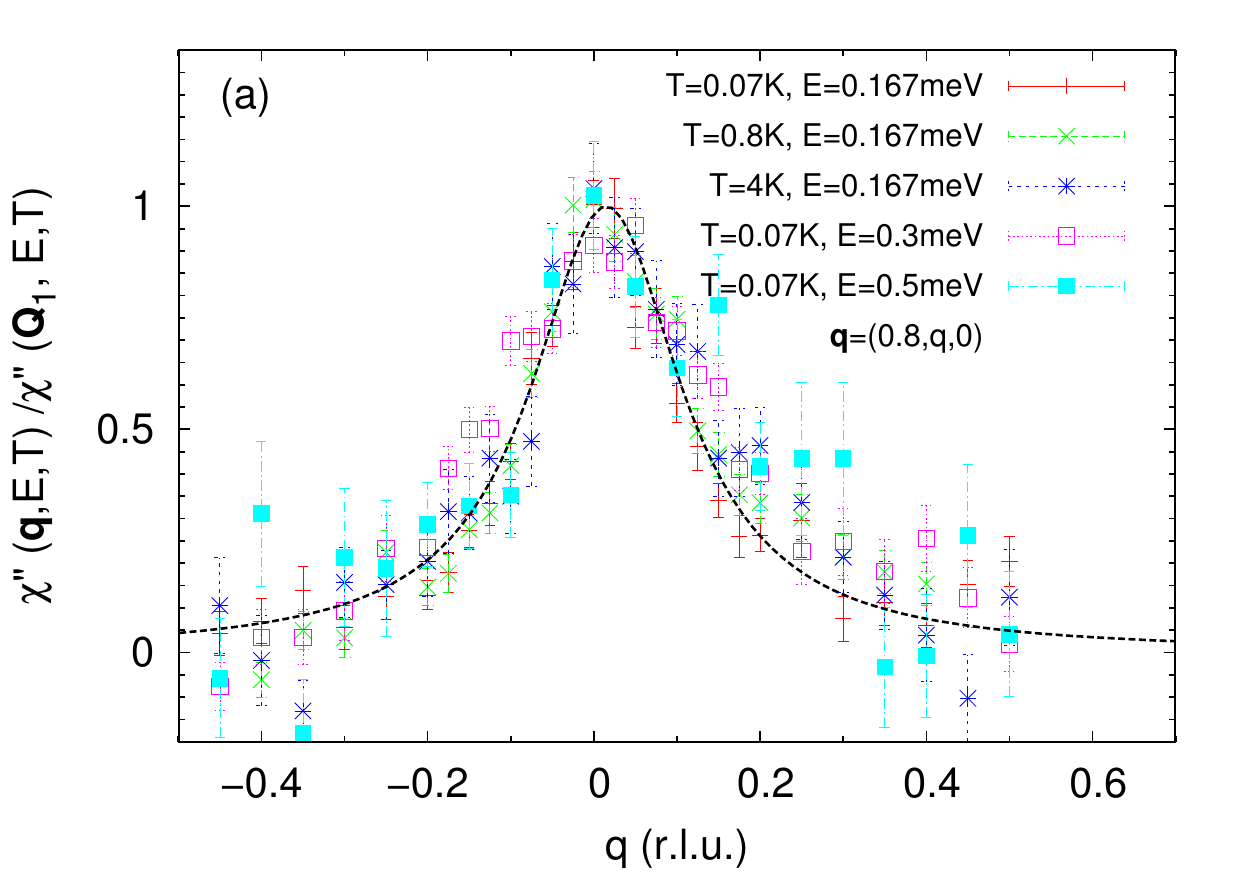}
\includegraphics[width=0.6\columnwidth]{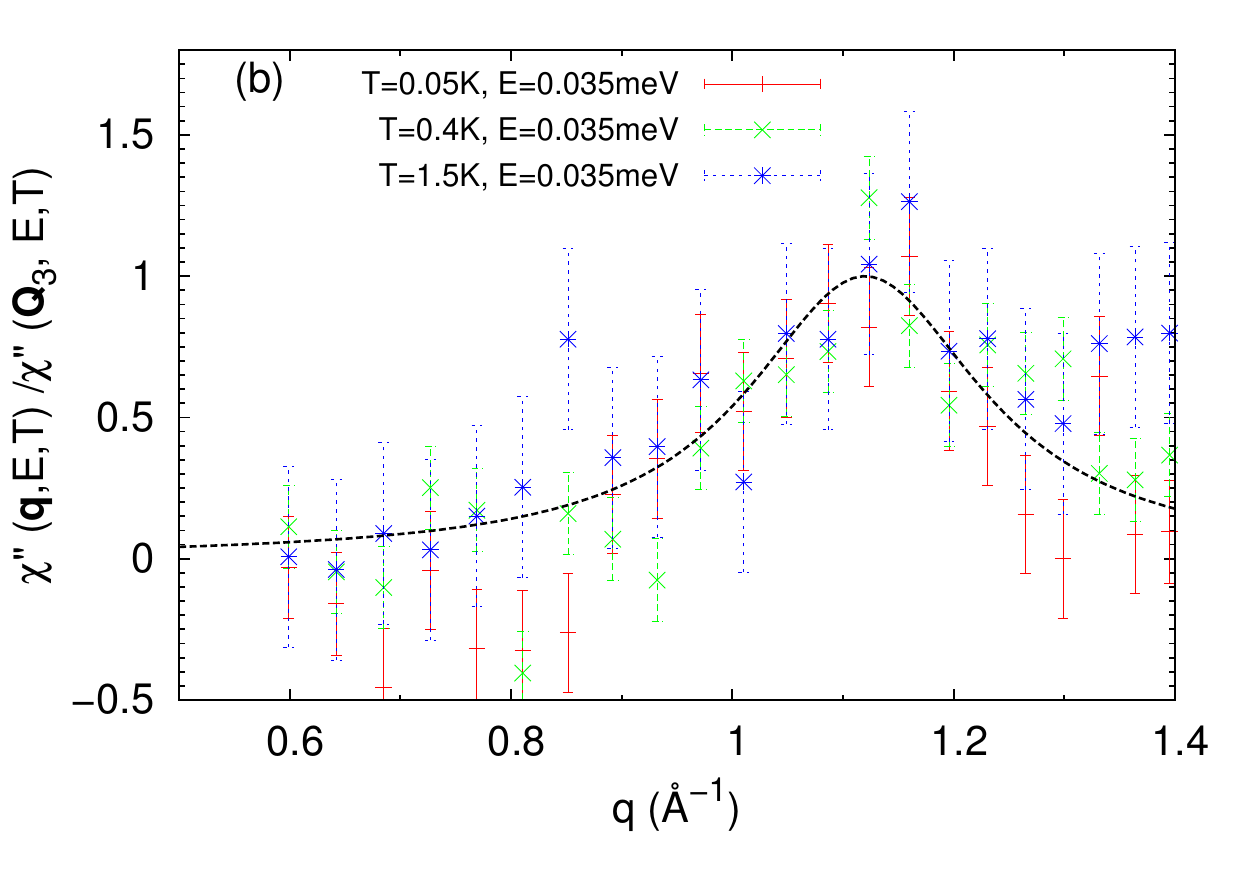}
\caption{$\chi '' ({\bf q}, E, T)$ as functions of ${\bf q}$ for two $q$-scans around ${\bf Q}_1$=(0.8,0,0) (a) and ${\bf Q}_3$=(1.4, 0, 0.3) (b) at various fixed $E$ and $T$ for CeCu$_{5.9}$Au$_{0.1}$. A q-independent background contribution has been subtracted. The fitting curve is Lorentzian $1/[(q-q_c)^2/\kappa_q^2+1]$ with $\kappa_q$=0.11 r.l.u. $\approx$ 0.13 \AA$^{-1}$ (considering $b=5.1$\AA) in (a) and $\kappa_q$=0.13 \AA$^{-1}$ in (b).  }
\label{Fig:cecuau_q}
\end{figure}

We first consider the ${\bf q}$-dependence of the measured susceptibility \cite{AS98} $\chi''({\bf q}, E, T)$ with the frequency and the temperature fixed near two ordering wavevectors ${\bf Q}_{1}$=(0.8,0,0) and ${\bf Q}_3$=(1.4,0,0.3). The ${\bf q}$-scan directions are also shown in Fig. \ref{Fig:cecuau_qorder}: along $b$-axis for ${\bf Q}_1$, and along a trajectory in ($ac$)-plane for ${\bf Q}_3$. We normalize the results for different frequencies and temperature to their peak value at ${\bf q} = {\bf Q}$. Such normalized plots as a function of the deviation of ${\bf q}$ from ${\bf Q}_1$ are shown in Fig. (\ref{Fig:cecuau_q}a).  A simple Lorentzian, as in Eq. (\ref{chi-tr}) fits the data, with a width  $\kappa_q \approx 0.13$\AA$^{-1}$, which is independent of temperature or frequency in the range measured and within the error bars of the data. This independence is a test of the separability of the ${\bf q}$ and $E$-dependence of the fluctuations. In  Fig. (\ref{Fig:cecuau_q}b), we show measurements near ${\bf Q}_3$ at different temperatures \cite{AS00}, which have much larger error bar (due to the low transfer energy). The best fit to the data shows the same width, independent of frequency and temperature as in Fig. (\ref{Fig:cecuau_q}a).

\begin{figure}[tbh]
\centering
\includegraphics[width=0.45\columnwidth]{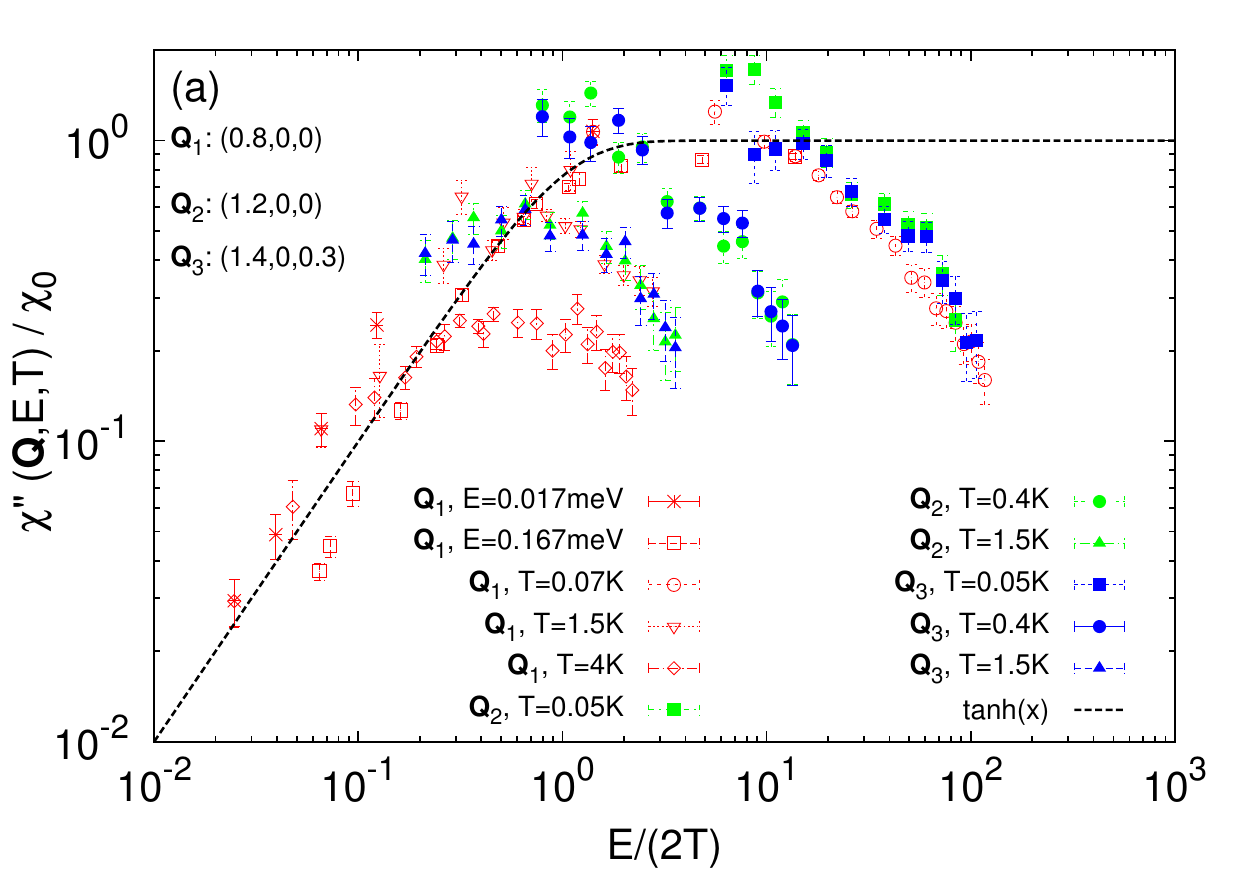}
\includegraphics[width=0.45\columnwidth]{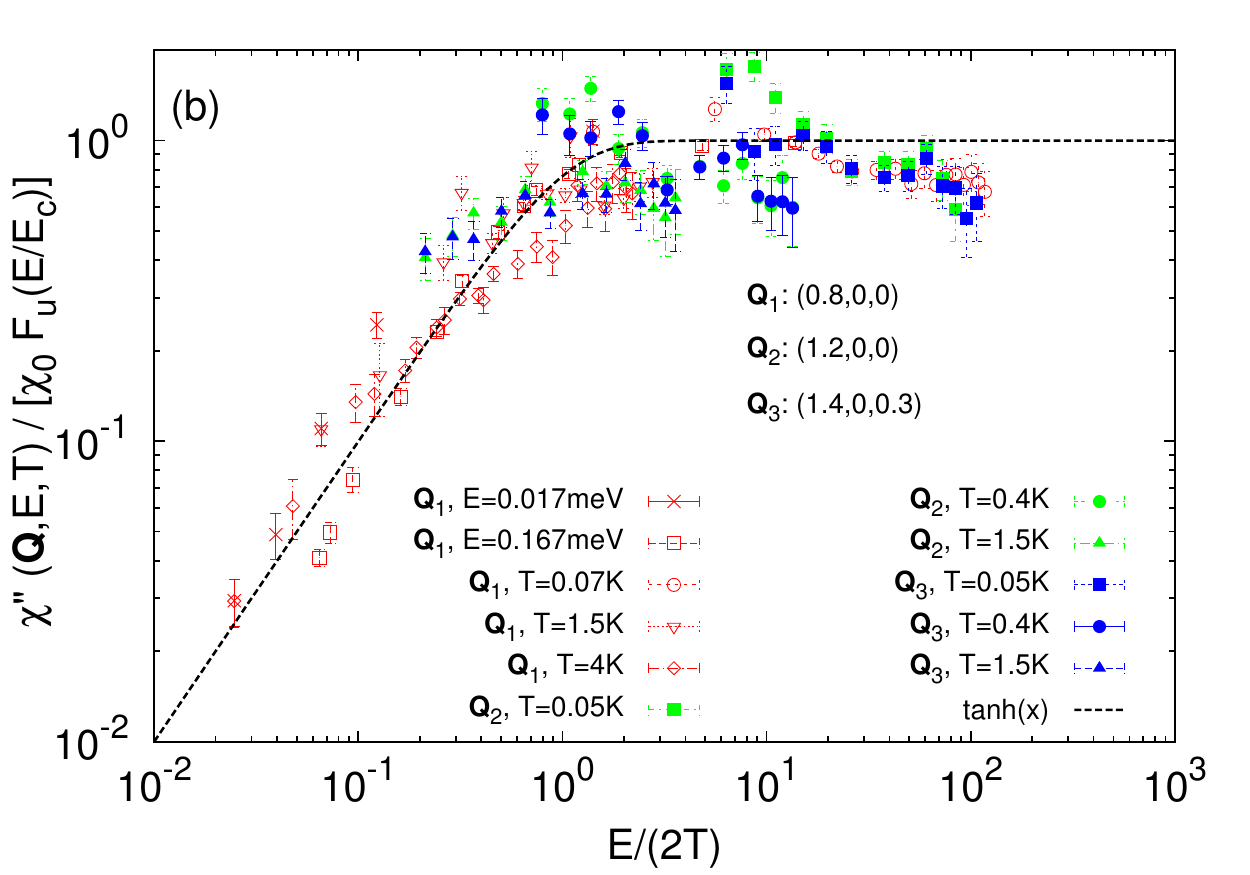}
\caption{ $\chi ''({\bf Q}, E, T)$ as functions of $E/(2T)$ for various constant-$E$ or $T$ scans for CeCu$_{5.9}$Au$_{0.1}$ at three ordering wave vectors  ${\bf Q}={\bf Q}_{1,2,3}$. The label for different symbols shows the location of ${\bf Q}$ as well as the fixed value of $E$ (or $T$), while $T$ (or $E$) is varied.  In (a),  $\chi ''({\bf Q}, E, T)$ is scaled by a constant $\chi_0 = 5.5 \mu_B^2(meV)^{-1}$ while in (b) it is scaled additionally by ${\mathcal F}_u(E/E_c) = 1/\sqrt{1+(E/E_c)^2}$ with $E_c=4K$. }
\label{fig:cecuau_et}
\end{figure}

We now turn to the frequency and temperature dependence at ${\bf q} ={\bf Q}_{1,3}$ as well as ${\bf Q}_2$=(1.2,0,0). From $\kappa_q \approx 0.13$ \AA$^{-1}$, we estimate that the deviation from criticality is very small for the sample measured, with $\sqrt{(p-p_c)/p_c} \approx 0.1$. Taking $E_c \approx 4 K$, as will be justified below, we find from Eq. (\ref{xi-tr}) that $\kappa_{E} \approx 4 \times e^{-10}$ K, which is much smaller than the range of temperature measurements. Therefore, one is well justified in comparing with the theory for the dependence on frequency and temperature at criticality, Eq. (\ref{chi-tr}), with $\xi_{\tau}^{-1} = 0$, where it should simply be proportional to $\tanh{(E/2T)}$. However, unlike the case of the Fe compound discussed above, the $E$ and $T$ of the measurements go well across the Fermi-energy of about 4 K, estimated from the linear part of the specific heat \cite{HvL96}. The cut-off function ${\mathcal{F}}_u(E/E_c)$ due to the upper cutoff $E_c$ can no longer be approximated as 1.  The measured \cite{AS98,AS00} $\chi '' ({\bf Q}, E, T)$, scaled by a constant $\chi_0$ [presumably the value of $\chi '' ({\bf Q}, E, T)$ at $E/(2T)\to\infty$ and $E \ll E_c$], is shown as a function of $E/2T$ in Fig.(\ref{fig:cecuau_et}a). The data agrees reasonably with the function $\tanh{(E/2T)}$  when $E \ll E_c$, but systematically deviates when $E \gtrsim E_c$ . We now choose a cut-off function 
\be
{\mathcal{F}}_u\left(\frac{E}{E_c}\right) = \frac{1}{\sqrt{1 + (E/E_c)^2}}.
\ee
We replot the same data, $\chi ''({\bf Q}, E, T)/\chi_0$, divided by ${\mathcal{F}}_u(E/E_c)$, as a function of $E/2T$ in Fig. (\ref{fig:cecuau_et}b). With a value of $E_c$ = 4K, the data, within its considerable error bars, is consistent with the scaling function $\tanh{(E/2T)}$. 

A complete test of the theory requires measurements varying $x$ or pressure to vary the distance to criticality and thereby test the theory of the correlation length. It would also be desirable to have more measurements for the momentum-energy and temperature dependence for smaller $E/T$. CeCu$_{6-x}$Au$_x$ has a rather complicated magnetic structure. We urge neutron scattering results also on other heavy-fermions with simpler antiferromagnetic structure near their quantum-criticality.

The ${\bf q}$ and $E$ dependence of the data also has been fitted to an alternate phenomenological form earlier\cite{AS00}. However, it does not give the observed linear in T resistivity. 

\section{Discussion}
The limited available data is consistent with the separability of the momentum and energy dependence of the critical fluctuations in two completely different experimental system, which share the feature that they both have itinerant AFM quantum critical points. The $\tanh{(E/2T)}$ dependence for energies smaller than the cut-off is also consistent with the data. As mentioned the linear in $T$-resistivity, and the $T \ln T$ entropy and thermopower are also properties of transport due to coupling of fermions to such fluctuations. Since an external field provides a linear change to the energy for spin-systems, one expects the resistivity to be a function of $(|H|/T)$. The resistance variation in a field at various temperatures has been found to be proportional to $\sqrt{\mu_B^2H^2 + T^2}$ \cite{Analytis}. Magnetic fluctuations of the form of Eq. (\ref{chi-tr}) provide a constant contribution to the nuclear relaxation rate. This has also been observed \cite{Zheng} near quantum-criticality.

Finally, we note that in cuprates for dopings near the quantum-critical region of the AFM but above a temperature, which appears to be determined by impurities, the resistivity is linear in T. In this region the spectral function of AFM-fluctuations, as determined by neutron scattering, does have $E/T$ scaling and a temperature independent correlation length. See Refs. \onlinecite{Hayden91, Keimer91}. 

There are many other materials showing quasi-2d fluctuations, and/or linear-$T$ resistivity, which could also be tested by the criticality of 2d quantum-dissipative XY model.
Heavy Fermions are notoriously hard to grow as large single crystals suitable for inelastic neutron scattering but the Ce compounds in the (115) family appear to be suitable for the purpose. Measurements have not been done on them in a suitable range of $({\bf q},E, T)$ to check the theory. The Fe superconductors appear also appear to be suitable for the purpose but systematic quantitative measurements of the kind needed are scarce. We urge such measurements.

{\it Acknowedgement} We wish to acknowledge discussions with Dr. D. M. Inosov on the measurements in the Fe-compound discussed in this paper. The work of CMV and LZ is supported partially by a National Science Foundation grant DMR 1206298.

%

\end{document}